\input harvmac

\noblackbox

\let\includefigures=\iftrue
\let\useblackboard=\iftrue
\newfam\black

\includefigures
\message{If you do not have epsf.tex (to include figures),}
\message{change the option at the top of the tex file.}
\input epsf
\def\figin{\epsfcheck\figin}\def\figins{\epsfcheck\figins}
\def\epsfcheck{\ifx\epsfbox\UnDeFiNeD
\message{(NO epsf.tex, FIGURES WILL BE IGNORED)}
\gdef\figin##1{\vskip2in}\gdef\figins##1{\hskip.5in}
\else\message{(FIGURES WILL BE INCLUDED)}%
\gdef\figin##1{##1}\gdef\figins##1{##1}\fi}
\def\DefWarn#1{}
\def\figinsert{\goodbreak\midinsert}
\def\ifig#1#2#3{\DefWarn#1\xdef#1{Fig.~\the\figno}
\writedef{#1\leftbracket Fig.\noexpand~\the\figno}%
\figinsert\figin{\centerline{#3}}\medskip\centerline{\vbox{
\baselineskip12pt\advance\hsize by -1truein
\noindent\footnotefont{\bf Fig.~\the\figno:} #2}}
\bigskip\endinsert\global\advance\figno by1}
\else
\def\ifig#1#2#3{\xdef#1{Fig.~\the\figno}
\writedef{#1\leftbracket Fig.\noexpand~\the\figno}%
\global\advance\figno by1} \fi

\def\doublefig#1#2#3#4{\DefWarn#1\xdef#1{Fig.~\the\figno}
\writedef{#1\leftbracket Fig.\noexpand~\the\figno}%
\figinsert\figin{\centerline{#3\hskip1.0cm#4}}\medskip\centerline{\vbox{
\baselineskip12pt\advance\hsize by -1truein
\noindent\footnotefont{\bf Fig.~\the\figno:} #2}}
\bigskip\endinsert\global\advance\figno by1}

\useblackboard
\message{If you do not have msbm (blackboard bold) fonts,}
\message{change the option at the top of the tex file.}
\font\blackboard=msbm10 scaled \magstep1 \font\blackboards=msbm7
\font\blackboardss=msbm5 \textfont\black=\blackboard
\scriptfont\black=\blackboards \scriptscriptfont\black=\blackboardss

\else

\fi
%

\def\yboxit#1#2{\vbox{\hrule height #1 \hbox{\vrule width #1
\vbox{#2}\vrule width #1 }\hrule height #1 }}
\def\fillbox#1{\hbox to #1{\vbox to #1{\vfil}\hfil}}
\def\ybox{{\lower 1.3pt \yboxit{0.4pt}{\fillbox{8pt}}\hskip-0.2pt}}
%
%


\def\comments#1{}



\def\II{\relax{I\kern-.10em I}}

\def\IZ{\relax{\rm Z\kern-.34em Z}}
\def\IB{\relax{\rm I\kern-.18em B}}
\def\IC{{\relax\hbox{$\inbar\kern-.3em{\rm C}$}}}
\def\ID{\relax{\rm I\kern-.18em D}}
\def\IE{\relax{\rm I\kern-.18em E}}
\def\IF{\relax{\rm I\kern-.18em F}}
\def\IG{\relax\hbox{$\inbar\kern-.3em{\rm G}$}}
\def\IGa{\relax\hbox{${\rm I}\kern-.18em\Gamma$}}
\def\IH{\relax{\rm I\kern-.18em H}}
\def\II{\relax{\rm I\kern-.18em I}}
\def\IK{\relax{\rm I\kern-.18em K}}
\def\IP{\relax{\rm I\kern-.18em P}}

%

\def\inbar{\,\vrule height1.5ex width.4pt depth0pt}

\def\IR{\relax{\rm I\kern-.18em R}}

\def\simgt{\hskip0.05in\relax{
\raise3.0pt\hbox{ $>$ {\lower5.0pt\hbox{\kern-1.05em $\sim$}} }}
\hskip0.05in}

%

\def\BP{\IP}

\def\lp10{\ell_p^{10}}
\def\lp11{\ell_p^{11}}
\def\R11{R_{11}}

\def\frac#1#2{{#1 \over #2}}



\newdimen\tableauside\tableauside=1.0ex
\newdimen\tableaurule\tableaurule=0.4pt
\newdimen\tableaustep
\def\phantomhrule#1{\hbox{\vbox to0pt{\hrule height\tableaurule width#1\vss}}}
\def\phantomvrule#1{\vbox{\hbox to0pt{\vrule width\tableaurule height#1\hss}}}
\def\sqr{\vbox{%
  \phantomhrule\tableaustep
  \hbox{\phantomvrule\tableaustep\kern\tableaustep\phantomvrule\tableaustep}%
  \hbox{\vbox{\phantomhrule\tableauside}\kern-\tableaurule}}}
\def\squares#1{\hbox{\count0=#1\noindent\loop\sqr
  \advance\count0 by-1 \ifnum\count0>0\repeat}}
\def\tableau#1{\vcenter{\offinterlineskip
  \tableaustep=\tableauside\advance\tableaustep by-\tableaurule
  \kern\normallineskip\hbox
    {\kern\normallineskip\vbox
      {\gettableau#1 0 }%
     \kern\normallineskip\kern\tableaurule}%
  \kern\normallineskip\kern\tableaurule}}
\def\gettableau#1 {\ifnum#1=0\let\next=\null\else
  \squares{#1}\let\next=\gettableau\fi\next}

\tableauside=1.0ex \tableaurule=0.4pt


 %
 %
 \def\eqnn#1{\xdef #1{(\secsym\the\meqno)}\writedef{#1\leftbracket#1}%
 \global\advance\meqno by1\wrlabeL#1}
 \def\eqna#1{\xdef #1##1{\hbox{$(\secsym\the\meqno##1)$}}
 \writedef{#1\numbersign1\leftbracket#1{\numbersign1}}%
 \global\advance\meqno by1\wrlabeL{#1$\{\}$}}
 \def\eqn#1#2{\xdef #1{(\secsym\the\meqno)}\writedef{#1\leftbracket#1}%
 \global\advance\meqno by1$$#2\eqno#1\eqlabeL#1$$}

\global\newcount\itemno \global\itemno=0

\def\itemaut#1{\global\advance\itemno by1\noindent\item{\the\itemno.}#1}



\hyphenation{Di-men-sion-al}



\lref\dinesymm{
  M.~Dine, Y.~Nir and Y.~Shadmi,
  ``Enhanced symmetries and the ground state of string theory,''
  Phys.\ Lett.\ B {\bf 438}, 61 (1998)
  [hep-th/9806124].
}

\lref\SchmaltzQS{
  M.~Schmaltz and R.~Sundrum,
  ``Conformal sequestering simplified,''
  hep-th/0608051.
}

\lref\Jay{N. Craig, P. Fox, J. Wacker, to appear.}

\lref\trapping{
  L.~Kofman, A.~Linde, X.~Liu, A.~Maloney, L.~McAllister and E.~Silverstein,
  ``Beauty is attractive: Moduli trapping at enhanced symmetry points,''
  JHEP {\bf 0405}, 030 (2004)
  [hep-th/0403001].
}

\lref\GRreview{
  G.~F.~Giudice and R.~Rattazzi,
  ``Theories with gauge-mediated supersymmetry breaking,''
  Phys.\ Rept.\  {\bf 322}, 419 (1999)
  [hep-ph/9801271].
}

\lref\SSreview{
  Y.~Shadmi and Y.~Shirman,
  ``Dynamical supersymmetry breaking,''
  Rev.\ Mod.\ Phys.\  {\bf 72}, 25 (2000)
  [hep-th/9907225].
}

\lref\Lreview{
  M.~A.~Luty,
  ``2004 TASI lectures on supersymmetry breaking,''
  hep-th/0509029.
}

\lref\wittenDSB{
  E.~Witten,
  ``Dynamical Breaking Of Supersymmetry,''
  Nucl.\ Phys.\ B {\bf 188}, 513 (1981).
}

\lref\earlyDSBmodels{
  M.~Dine, A.~E.~Nelson and Y.~Shirman,
  ``Low-Energy Dynamical Supersymmetry Breaking Simplified,''
  Phys.\ Rev.\ D {\bf 51}, 1362 (1995)
  [hep-ph/9408384];
  M.~Dine, A.~E.~Nelson, Y.~Nir and Y.~Shirman,
  ``New tools for low-energy dynamical supersymmetry breaking,''
  Phys.\ Rev.\ D {\bf 53}, 2658 (1996)
  [hep-ph/9507378].
}

\lref\progress{
  E.~Poppitz and S.~P.~Trivedi,
  ``New models of gauge and gravity mediated supersymmetry breaking,''
  Phys.\ Rev.\ D {\bf 55}, 5508 (1997)
  [hep-ph/9609529];
  N.~Arkani-Hamed, J.~March-Russell and H.~Murayama,
   ``Building models of gauge-mediated supersymmetry breaking without a
  messenger sector,''
  Nucl.\ Phys.\ B {\bf 509}, 3 (1998)
  [hep-ph/9701286];
}

\lref\savasI{
  S.~Dimopoulos, G.~R.~Dvali, R.~Rattazzi and G.~F.~Giudice,
  ``Dynamical soft terms with unbroken supersymmetry,''
  Nucl.\ Phys.\ B {\bf 510}, 12 (1998)
  [hep-ph/9705307].
}

\lref\binetruy{
  P.~Binetruy and E.~Dudas,
  ``Gaugino condensation and the anomalous U(1),''
  Phys.\ Lett.\ B {\bf 389}, 503 (1996)
  [hep-th/9607172].
}

\lref\dvali{
  G.~R.~Dvali and A.~Pomarol,
   ``Anomalous U(1), gauge-mediated supersymmetry breaking and Higgs as
  pseudo-Goldstone bosons,''
  Nucl.\ Phys.\ B {\bf 522}, 3 (1998)
  [hep-ph/9708364].
}

\lref\izawa{
  K.~I.~Izawa, Y.~Nomura, K.~Tobe and T.~Yanagida,
  ``Direct-transmission models of dynamical supersymmetry breaking,''
  Phys.\ Rev.\ D {\bf 56}, 2886 (1997)
  [hep-ph/9705228].
}

\lref\murayama{
  H.~Murayama,
  ``A model of direct gauge mediation,''
  Phys.\ Rev.\ Lett.\  {\bf 79}, 18 (1997)
  [hep-ph/9705271].
}

\lref\shirman{ Y.~Shirman,
   ``New models of gauge mediated dynamical supersymmetry breaking,''
   Phys.\ Lett.\ B {\bf 417}, 281 (1998)
   [hep-ph/9709383].
}

\lref\luty{
  M.~A.~Luty,
  ``Simple gauge-mediated models with local minima,''
  Phys.\ Lett.\ B {\bf 414}, 71 (1997)
  [hep-ph/9706554].
}

\lref\savas{
  S.~Dimopoulos, G.~R.~Dvali and R.~Rattazzi,
   ``A Simple Complete Model Of Gauge-Mediated Susy-Breaking And Dynamical
  Relaxation Mechanism For Solving The Mu Problem,''
  Phys.\ Lett.\ B {\bf 413}, 336 (1997)
  [hep-ph/9707537].
}

\lref\agashe{
  K.~Agashe,
  ``An Improved Model Of Direct Gauge Mediation,''
  Phys.\ Lett.\ B {\bf 435}, 83 (1998)
  [hep-ph/9804450].
}

\lref\BT{
  J.~D.~Brown and C.~Teitelboim,
  ``Neutralization of the cosmological constant by membrane creation,''
  Nucl.\ Phys.\ B {\bf 297}, 787 (1988).
  }

\lref\BP{ R.~Bousso and J.~Polchinski,
   ``Quantization of four-form fluxes and dynamical neutralization of the
  cosmological constant,''
  JHEP {\bf 0006}, 006 (2000)
  [hep-th/0004134].
}

\lref\FMSW{
  J.~L.~Feng, J.~March-Russell, S.~Sethi and F.~Wilczek,
  ``Saltatory relaxation of the cosmological constant,''
  Nucl.\ Phys.\ B {\bf 602}, 307 (2001)
  [hep-th/0005276].
}

\lref\land{
  E.~Silverstein,
  ``(A)dS backgrounds from asymmetric orientifolds,''
  contribution to Strings 2001 [hep-th/0106209];
  A.~Maloney, E.~Silverstein and A.~Strominger,
  ``De Sitter space in noncritical string theory,''
  in {\it Cambridge 2002: The future of theoretical physics and
  cosmology}, 570-591
 [hep-th/0205316];
  S.~Kachru, R.~Kallosh, A.~Linde and S.~P.~Trivedi,
  ``De Sitter Vacua In String Theory,''
  Phys.\ Rev.\ D {\bf 68}, 046005 (2003)
  [hep-th/0301240].
}

\lref\KPV{
  S.~Kachru, J.~Pearson and H.~L.~Verlinde,
   ``Brane/flux annihilation and the string dual of a non-supersymmetric  field
  theory,''
  JHEP {\bf 0206}, 021 (2002)
  [hep-th/0112197].
}

\lref\ISnotes{
  K.~A.~Intriligator and N.~Seiberg,
  ``Lectures on supersymmetric gauge theories and electric-magnetic  duality,''
  Nucl.\ Phys.\ Proc.\ Suppl.\  {\bf 45BC}, 1 (1996)
  [hep-th/9509066].
}

\lref\DF{
  M.~Dine and W.~Fischler,
  ``A Phenomenological Model Of Particle Physics Based On Supersymmetry,''
  Phys.\ Lett.\ B {\bf 110}, 227 (1982).
}

\lref\ISS{
  K.~Intriligator, N.~Seiberg and D.~Shih,
  ``Dynamical SUSY breaking in meta-stable vacua,''
  JHEP {\bf 0604}, 021 (2006)
  [hep-th/0602239].
}

\Title{\vbox{\baselineskip12pt\hbox{}
\hbox{SU-ITP-06/22}\hbox{SLAC-PUB-12059}\hbox{UCI-TR-2006-14 }} }
{\vbox{ \centerline{Retrofitting O'Raifeartaigh Models}
\centerline{with Dynamical Scales}
}}
\bigskip
\bigskip
\centerline{Michael Dine$^1$, Jonathan L. Feng$^2$, and Eva
Silverstein$^3$}
\bigskip
\centerline{$^1$ {\it Santa Cruz Institute for Particle Physics,
Santa Cruz CA 95064, USA}} \centerline{$^2${\it Department of
Physics and Astronomy, University of California, Irvine, CA 92697,
USA} } \centerline{$^3${\it SLAC and Department of Physics, Stanford
University, Stanford, CA 94305-4060, USA}}
\bigskip
\bigskip
\noindent

We provide a method for obtaining simple models of supersymmetry
breaking, with all small mass scales generated dynamically, and
illustrate it with explicit examples. We start from models of
perturbative supersymmetry breaking, such as O'Raifeartaigh and Fayet
models, that would respect an $R$ symmetry if their small input
parameters transformed as the superpotential does. By coupling the
system to a pure supersymmetric Yang-Mills theory (or a more general
supersymmetric gauge theory with dynamically small vacuum expectation
values), these parameters are replaced by powers of its dynamical
scale in a way that is naturally enforced by the symmetry. We show
that supersymmetry breaking in these models may be straightforwardly
mediated to the supersymmetric Standard Model, obtain complete models
of direct gauge mediation, and comment on related model building
strategies that arise in this simple framework.

\bigskip
\Date{August 2006}

\newsec{Introduction and General Idea}

Dynamical supersymmetry (SUSY) breaking \wittenDSB\ and the
mediation of SUSY breaking to the supersymmetric Standard Model
(SSM) have been studied extensively. A particularly important
objective is to identify simple models of dynamical SUSY breaking
that may be straightforwardly mediated to the SSM, yielding
predictive and phenomenologically attractive superpartner spectra
\refs{\GRreview,\SSreview,\Lreview}. In early examples with
gauge-mediated SUSY breaking \earlyDSBmodels, the problems of SUSY
breaking and its mediation were addressed by postulating separate
SUSY breaking and messenger sectors.  These models motivated many
advances \refs{\progress,\savasI,\murayama}, culminating in a few
genuinely simple and viable models of direct gauge mediation
\refs{\binetruy,\izawa,\luty,\savas,\dvali,\shirman,\agashe}, in
which fields of the SUSY breaking sector also play the role of
messengers, transmitting SUSY breaking to the SSM.

In this note, we develop a straightforward method for obtaining
simple models of SUSY breaking in which all small scales are
generated dynamically.  We show further that SUSY breaking in these
models may be rather simply communicated to the SSM, providing new
avenues for direct gauge mediation and gravity mediation.  To
illustrate the method, we work through two complete gauge mediation
examples that are representative of large classes of models, and we
discuss the method's application to more general model building
problems.

The basic strategy in its simplest realization can be summarized as
follows:

\noindent (1) Start with a model of perturbative SUSY breaking, such
as an O'Raifeartaigh or Fayet model, whose small input parameters
$m_i$ break an $R$ symmetry that would be restored if the $m_i$
transformed as the superpotential does.

\noindent (2) Couple the system to a SUSY preserving sector with a
dynamically small operator vacuum expectation value (VEV). Our
prototypical example will be pure $SU(2)$ Yang-Mills theory, with
gauge field strength superfield $W_\alpha$ and dynamical scale
$\Lambda$. Replace dimensional parameters $m_i$ in the
superpotential by $W_\alpha W^\alpha$ suppressed by appropriate
powers of a high scale $M_*$. At low energies, $W_\alpha
W^\alpha\sim \Lambda^3$.  This renders the $m_i$ dynamically small
in a way naturally enforced by the symmetries and preserves a local
SUSY breaking minimum.

We will refer to this procedure (1)-(2) as {\it retrofitting} the
old-fashioned perturbative SUSY breaking models. Elementary
ingredients suffice to bring such models up to modern model building
standards of naturalness, while preserving some of the simplicity of
early constructions \DF.  In effect, we consider a {\it
supersymmetric} hidden sector to obtain dynamically small scales,
which allows the SUSY {\it breaking} sector to be more
directly coupled to the SSM.

If desired, the couplings to $W_\alpha W^\alpha$ can arise from purely
renormalizable interactions by integrating out massive flavors in the
$SU(2)$ SUSY gauge theory \ISnotes. In any case, the coupling to the
$SU(2)$ sector does not destroy the local SUSY breaking minimum of the
perturbative model (1), though it often introduces SUSY vacua far away
in field space. As discussed, for example, in
\refs{\murayama,\luty,\savas,\land,\KPV,\ISS}, we need not impose that
the SUSY breaking configuration be the global minimum of the
potential.

Indeed, one element of many successful models is metastability.  In
field theory models of dynamical SUSY breaking, the requirement that
SUSY be broken in the global minimum is very restrictive, and
allowing for metastable vacua greatly simplifies the problems of
model building, especially for gauge mediation. This point was
emphasized clearly, for example, in
\refs{\murayama,\savasI,\luty,\savas}; more recently it has found
application in the problem of moduli stabilization \land\ and
dynamics \refs{\BT,\BP,\FMSW}, in the vacuum structure of large $N$
gauge theories arising in generalizations of AdS/CFT \KPV, and in
supersymmetric QCD \ISS.

As we will see, simple constructions lead almost trivially to a
large class of dynamical SUSY breaking models and suggest an array
of further model building possibilities. It is worth remarking that
the models need not be chiral and can have non-vanishing Witten
index, like the models of \refs{\KPV,\ISS}. They can possess
interesting (discrete) symmetries, which naturally protect the
structures required for model building goals. This simple method
allows construction of theories with direct gauge mediation as well
as gravity-mediated models with appropriately large
(non-loop-suppressed) gaugino masses.

As discussed recently in \ISS, some basic classes of supersymmetric
gauge theories reduce at low energies to infrared-free O'Raifeartaigh
models with metastable SUSY breaking.  In some circumstances,
the direct mediation models of the type we consider here may be UV
completed by asymptotically-free quantum field theory.  In other
circumstances, the models may be completed by string theory, where
metastable SUSY breaking \land\ has played a crucial role.

{}From the perspective of weakly coupled string theory, one might
worry that there are additional approximate moduli that affect the
value of the gauge coupling.  It is worth noting in this connection
that the current state of the art in string moduli stabilization ---
via a combination of a tree level potential, orientifolds, and
Ramond-Ramond fluxes --- can fix the dilaton and other moduli at a
high scale. In the context of low energy supersymmetric models, this
allows for a gaugino condensate which does not vary with extra moduli
beyond those evident in the low energy field theory of interest here,
whose couplings are fixed by discrete symmetries.

In the next section, we consider retrofitting a class of
O'Raifeartaigh models and work through a simple example in detail.  We
next simplify the model further to extract some lessons about the role
of chirality and symmetry.  We follow this in \S3 with another general
class of models including a Fayet-Iliopoulos parameter. In the final
section, we summarize and discuss further model building applications.

\newsec{Retrofitting O'Raifeartaigh Models}

In this section we will implement the procedure outlined above in
concrete examples and comment on model building lessons that arise in
this framework. We begin with a brief review of O'Raifeartaigh models
and their challenges. Next we consider a simple explicit example which
we retrofit to render its scales dynamical in a way consistent with
symmetries.  This model is complete in that it readily incorporates
messengers appropriate for gauge mediation, generating Standard Model
superpartner masses.  In the final subsection we extract lessons
illustrated by even simpler systems, emphasizing the role of
metastability in avoiding the unnecessary constraints of chirality and
vanishing Witten index.

Consider O'Raifeartaigh models, with $n$ fields $Z_1,\dots,Z_n$, $n'$
fields $\phi_1,\dots,\phi_{n'}$, $n'<n$, and superpotential
\eqn\ORgen{W=\sum_i Z_i f_i(\phi_a) \ .}
This class of models breaks SUSY classically for generic choices of
functions $f$. At tree level, its main shortcomings are (i) there is
automatically a flat direction in its potential, (ii) it does not
automatically provide messengers and $R$ symmetry breaking as
required to mediate SUSY breaking to the SSM, and (iii) its scales
are input by hand, with some couplings set to zero without a
symmetry reason. (With regard to point (ii), for definiteness we
here consider gauge-mediated SUSY breaking, and consider more
general applications in the later discussion.)

We will address each of these, illustrating the technique with perhaps
the simplest version of \ORgen.  Let us first summarize the method.
With regard to point (i), the Coleman-Weinberg potential expanded
about an appropriate point in field space generically lifts the flat
direction; one can explicitly check for self consistent metastable
solutions as in \refs{\DF,\savas}. Point (ii) can be addressed by
coupling in messengers and including their contribution to the
Coleman-Weinberg potential self-consistently.  Finally point (iii) can
be addressed by coupling in an otherwise supersymmetric $SU(2)$
gaugino condensate, or any other more general SUSY sector with a
dynamically small operator VEV.

\subsec{A Complete, Simple Example}

As a very simple illustrative example, consider a model with
messengers $\eta$ and $\tilde\eta$ in, say, the ${\bf 5}$ and ${\bf
\bar 5}$ of $SU(5)$, and three fields $Z_1$, $Z_2$, and $\phi$.
A natural superpotential based on the O'Raifeartaigh paradigm \ORgen\
is
\eqn\ORsimple{W=Z_1{\phi^3\over{3 M_*}}+Z_2
\bigg(\lambda{\phi^2\over 2}[1+\lambda_1{Z_2\over
M_*}]-\lambda{\mu^2\over 2}+{{\phi\eta\tilde\eta}\over{M_*}}
\bigg)+\lambda\phi\eta\tilde\eta +\lambda_2{(\eta\tilde\eta)^2\over
M_*} \ , }
where $M_*$ is a high scale corresponding to new(er) physics, such
as a grand unified or Kaluza-Klein scale.  We will obtain the
parameter $\mu^2 \sim \Lambda^3/M_*$ dynamically from a coupling
$\int d^2\theta W_\alpha W^\alpha {\lambda Z_2\over M_*}$ between
the $SU(2)$ sector and the O'Raifeartaigh model.

This theory is invariant under the following two symmetries: a
discrete $\IZ_{2N}$ $R$ symmetry, with $N>2$, under which the
superpotential transforms with charge 2 and the fields $\phi$,
$Z_1$, $Z_2$, $W_\alpha$, and $\eta\tilde\eta$ have charges 1, $-1$,
0, 1, and 1; and a continuous $R$ symmetry, under which $\phi$ is
neutral and $Z_1$, $Z_2$, and $\eta\tilde\eta$ transform, which
governs the renormalizable terms, but is broken by the
$M_*$-suppressed operators. The superpotential of \ORsimple\ is the
most general one respecting these symmetries, up to terms higher
order in $M_*^{-1}$.

In the absence of the messengers, the model has a massless
combination of $Z_1$ and $Z_2$, and a $\phi$ VEV
\eqn\phiVEV{\phi_0^2\approx
{\mu^2\over{1+2\lambda_1Z_2/M_*}}-{2\mu^4\over
{3M_*^2[1+2\lambda_1Z_2/M_*]^4\lambda^2}}}
Plugging in this solution yields $F_{Z_{1,2}}$ terms of order
\eqn\Ftwo{  F_{Z_1} \approx {\mu^3 \over 3 M_*} \ , \quad F_{Z_2}
\approx -{\mu^4\over 3 M_*^2 \lambda} \ . }
plus corrections down by powers of $\mu/M_*$ and $Z_2/M_*$. In the
full model, $F_{Z_2}$ couples to the messengers $\eta,\tilde\eta$,
suppressed by an additional power of $\phi_0/M_*\approx \mu/M_*$.
(This suppression is forced on the model by the discrete symmetries
it respects.)  An $F$ term for $Z$ combined with a VEV for $\phi$
will produce naturally small superpartner masses as we will discuss
further below.

The Coleman-Weinberg potential obtained by integrating out
$\phi,\eta,\tilde\eta$ yields a metastable minimum for $Z$ at the
origin in a self-consistent expansion about $\phi=\phi_0,
\eta=\tilde\eta=0$. Let us begin by integrating out the fluctuations
of $\phi$; we will show that these dominate over messenger loops in
this model.\foot{In the model of \S3, the messengers themselves will
play a leading role in stabilizing the scalar fields, providing a
particularly direct mediation mechanism.} Writing
$\phi=\phi_0+\delta\phi$, the mass terms for the fluctuations
$\delta\phi\equiv \delta\phi_1+i\delta\phi_2$ are of the form
\eqn\massphi{\lambda^2\delta\phi\delta\bar\phi  (
\mu^2+|Z_2|^2)+(\delta\phi_1^2-\delta\phi_2^2){\mu^4\over {3 M_*^2}}
\ , }
plus contributions subleading in the regime $\mu/M_*\ll 1$.  The
fermion loops cancel the $\delta\phi\delta\bar\phi$ contribution here,
and so the leading contribution to the potential from the $\phi$
multiplet is
\eqn\VdeltOR{\eqalign{ \Delta V &\approx {\rm Tr~ log}
\bigg[(\mu^2\lambda^2+|Z_2|^2\lambda^2+p^2)^2-\bigg({\mu^4\over{3
M_*^2}}\bigg)^2\bigg] -{\rm Tr~
log}[(\mu^2\lambda^2+|Z_2|^2\lambda^2+p^2)^2] \cr &= {1\over
32\pi^2}\bigg({\mu^4\over{3 M_*^2}}\bigg)^2{\rm
log}[\lambda^2(\mu^2+|Z_2|^2)/M_*^2] \ , \cr } }
plus subleading contributions.  The messenger loops are subleading
relative to the $\phi$ loops; the $\eta,\tilde\eta$ mass terms are
of the form $(|\eta|^2+|\tilde\eta|^2)\mu^2|\lambda+Z_2/M_*|^2$ plus
a SUSY breaking term proportional to $\eta\tilde\eta
\mu^5/(M_*^3\lambda)$, which will be much smaller than that in
\massphi.

Although subleading in the Coleman-Weinberg potential, the
messengers provide the dominant transmission of SUSY breaking to the
SSM.  As we just noted, the leading contribution to the messenger
masses is from the supersymmetric $\lambda\phi\eta\tilde\eta$
coupling, giving $m_{\eta,\tilde\eta}\sim \lambda\mu$, while the
leading SUSY breaking contribution to their masses is $\Delta
m^2_{\eta\tilde\eta}\sim \mu^5/(M_*^3\lambda)$.  In application to
gauge mediation, this yields gaugino and squark masses of the order
of
\eqn\gaugmass{\tilde m \sim {g^2\over 16 \pi^2} {\mu^4\over
M_*^3\lambda^2} \ ,}
where $g$ represents SSM gauge couplings.

For $F_{Z_i}\le 10^{20}~{\rm GeV}^2$, the gravity mediated
contribution to superpartner masses is suppressed relative to the
gauge mediated contribution.  Imposing this, we find that, for
example, $m_{\eta,\tilde\eta} \sim \lambda \mu \sim 10^{11}~{\rm GeV}$
and $M_*\sim 10^{15}~{\rm GeV}$ produces a viable model, with
$\lambda\sim 0.1$.  Of course, if $M_*$ were much lower than the GUT
scale, then the messenger scale could be lower as well.  As we will
discuss further in the next subsection, another application of our
method is to models where gravity mediation dominates.

To retrofit the model, as discussed above, we couple in a pure SUSY
Yang-Mills sector with gauge superfield $W_\alpha$, replacing the
superpotential \ORsimple\ with
\eqn\ORretro{W=Z_1{\phi^3\over{3 M_*}}+\bigg(-{1 \over 4 g^2}-
{\lambda Z_2 \over M_*}\bigg) W_\alpha^2 + Z_2
\bigg({\lambda\phi^2\over
2}[1+\lambda_1Z_2/M_*]+{{\phi\eta\tilde\eta}\over{M_*}}
\bigg)+\lambda\phi\eta\tilde\eta +{\lambda_2(\eta\tilde\eta)^2\over
M_*} \ . }
Integrating out the gauge interactions yields
\eqn\ORretroLam{W=Z_1{\phi^3\over{3 M_*}}+\lambda\Lambda^3 e^{-12
Z_2/b_0 M_*}+Z_2 \bigg(\lambda{\phi^2\over
2}[1+\lambda_1Z_2/M_*]+{{\phi\eta\tilde\eta}\over{M_*}} \bigg) +
\lambda\phi\eta\tilde\eta +{\lambda_2(\eta\tilde\eta)^2\over M_*} \
. }
Expanding in $Z_2$ yields at leading order a model of the form
\ORsimple, with $\mu^2\propto \Lambda^3/M_*$.  It is self-consistent
to integrate out the gauge degrees of freedom because they have
$M_*$-suppressed couplings to the rest of the system, too weak to
compete against the forces in the Yang-Mills sector proper, which
appear at the scale $\Lambda$.

Including the $Z_2$ dependence in solving for $\phi_0$ yields
\eqn\FZ{F_{Z_2}\propto {\Lambda^6\over M_*^4\lambda} e^{-24 Z_2/b_0
M_*}+\dots \ , }
generalizing \Ftwo.  This $Z_2$-dependence can lead to the presence
of supersymmetric minima far away for appropriate ranges of
parameters, but it does not destabilize our local minimum, as we can
see easily as follows. Expanding in $Z_2$, the term $|F_{Z_2}|^2$ in
the effective potential produces a tadpole of order $Z_2
\Lambda^{12}/(M_*^9\lambda^2)$.  The Coleman-Weinberg potential
\VdeltOR\ produces a mass term of order $\lambda^2|Z_2|^2
\Lambda^9/M_*^7$, sufficient to stabilize $Z_2$ close to its
original minimum at the origin.

\subsec{Remarks on Retrofitting O'Raifeartaigh Models}

In the previous subsection, we implemented the retrofitting procedure
in a complete model, which was natural, given the specified
symmetries, and incorporated messengers generating sparticle
masses. The method has wider applicability, and it is interesting to
extract and separate some of the essential elements of the procedure
and consider independently the role of symmetry, chirality (or lack
thereof), and metastability.

A simple example illustrates some of the main points.  Consider a
model with singlets $Z$, $A$ and $B$, and superpotential
\eqn\ORsimsup{W = MAB + \lambda Z(A^2-\mu^2) \ .}
This model breaks supersymmetry. For $M>\sqrt{2}\lambda \mu$, there is
a minimum in the $A$ direction at $\langle A \rangle =0$. At the
classical level, there is a flat direction; the expectation value of
$Z$ is undetermined. However, at one loop, the standard
Coleman-Weinberg calculation gives $\langle Z \rangle =0$. The
potential grows quadratically near the origin and logarithmically for
$Z \gg M$.

Before rendering the mass parameters dynamical, note that a small
deformation of the model makes the SUSY breaking minimum merely
metastable. If we write
\eqn\supcoup{W = MAB + \lambda Z(A^2-\mu^2)+ \epsilon M Z^2 \ , }
for sufficiently small $\epsilon$, there is still a metastable
minimum near the origin.  There is also a global SUSY preserving
minimum at $Z =\lambda \mu^2/(2\epsilon M)$.  (One can check that
there is still a massless goldstino in the metastable minimum.)

Now we can retrofit the model and generate the small parameters
dynamically.  First, replace the $\mu^2$ coupling by a coupling of $Z$
to a strongly interacting gauge theory. This can be simply a pure
supersymmetric gauge theory, leading to
\eqn\Wgauge{W = MAB + \lambda Z A^2 + \bigg(
-{1 \over 4 g^2}+{Z  \over M^*} \bigg) W_\alpha W^{\alpha} \ .}
We are assuming $g$ is fixed.  If there are other moduli-like fields
contributing to the gauge coupling, we assume that they are fixed at a
higher scale, e.g., by fluxes or other dynamics. Now $\mu^2$ is
related to the dynamical scale of the hidden sector theory;
integrating out the gauge interactions, the superpotential is
\eqn\supnext{W = MAB + \lambda Z A^2 + {\Lambda^3} e^{12 Z/b_0 M_*}\ .}
Expanding the exponential in powers of $Z$, the linear term
reproduces the original O'Raifeartaigh model.  Near the origin, the
Coleman-Weinberg corrections still generate a positive curvature.
This still leads to a local minimum, provided $M\ll M_*$. As $Z
\rightarrow -\infty$ (with $A=B=0$), the energy tends to zero and
SUSY is restored, though other effects may come in depending on the
UV completion of the system.

This model closely parallels the O'Raifeartaigh models arising in
the low energy limit of certain SUSY QCD theories \ISS\ in a number
of ways. With the small mass term for $Z$, $M\epsilon Z^2$, if
$M\epsilon$ is sufficiently small, there is still a local minimum
near the origin, but there is a supersymmetric minimum for $Z \sim
\lambda\mu^2/(2 M\epsilon)$. If the gauge group of the strongly
interacting sector is $SU(N)$, the index can be computed for
non-zero $\epsilon$, and it is equal to $N$. The analogous
statements hold for the models of \ISS\ for small quark mass.

With $\epsilon=0$, this model is the most general consistent with a
discrete $\IZ_{2N}$ $R$ symmetry, under which the fields $Z$, $A$, and
$B$, have charges 0, 1, and 1, respectively. The low energy theory has
an {\it approximate, continuous} $R$ symmetry under which $Z$, $A$,
and $B$ have charges 2, 0, and 2.

So far, this model has an additional scale $M$.  But we can make this
scale dynamical as well, without introducing any new scales beyond
$M^*$ and $\Lambda$.  Simply introduce two other singlets, $\chi$ and
$C$, with couplings
\eqn\moreflds{W_\chi = C AB+ \lambda \chi C^2 + a{\chi \over M^*}
W_\alpha W^{\alpha} \ .}
The parameter $a$ is naturally of order one, if $\chi$ is neutral
under the discrete $R$ symmetry. In contrast to our complete models in
\S2.1 and \S3, this structure is not enforced by symmetries, but it is
meant only to be illustrative.  The addition of small,
symmetry-preserving couplings does not alter its basic features.

All of this illustrates that it is easy to construct metastable models
of dynamical SUSY breaking with non-vanishing Witten index, which are
not (necessarily) chiral. (These features also appear in the
O'Raifeartaigh models in the infrared limit of some recent SUSY QCD
examples \ISS\ and earlier models of gauge-mediated SUSY breaking.)

Unlike in our complete example of \S2.1, in this case we did not
include messengers for gauge mediation.  A simple coupling
$Z\phi\tilde\phi$ would not suffice here since
$Z\sim\Lambda^3/M_*^2$ is extremely small in the minimum obtained
above (including the small tadpole introduced by the $Z$-dependence
of the $SU(2)$ gauge coupling). In \S2.1, we solved this problem via
a natural superpotential leading to spontaneous breaking of a
discrete $R$ symmetry.  That example was limited to high or
intermediate scale messenger masses, and it is of interest to
explore this vast class of retrofit O'Raifeartaigh models in search
of models with lower mass messengers.  In models of this type, the
SUSY breaking scale would be arbitrary.  If the approximate $R$
symmetries are broken at a scale of order the SUSY breaking scale,
then the messenger mass scale is arbitrary as well. This may allow
the construction of gauge-mediated models with scales of SUSY
breaking as low as $10~{\rm TeV}$.

On the other hand, it is also a very simple to consider these theories
as hidden sectors for gravity mediation.  These models are promising
from this viewpoint since no symmetry forbids a coupling of $Z$ to the
SSM gauginos.  In this case, the scalar and gaugino masses are of the
same order, rather than being suppressed by a loop factor, as in
anomaly mediation.

\newsec{Retrofitting Fayet Models}

Another class of illustrative examples includes Fayet models, another
of the classic models of perturbative SUSY breaking. We will start by
describing a version with two input parameters, at least one of which
needs to be small for natural SUSY breaking.  We then upgrade the
model to obtain the necessary small scale dynamically.  This class of
examples has the feature that the fields generating the leading
contributions to the Coleman-Weinberg potential also can play the role
of messengers of gauge-mediated SUSY breaking.

\subsec{The Perturbative SUSY Breaking Model}

Begin with gauge group $U(1)$ and chiral fields $X$, $\phi$, and
$\tilde\phi$ with charges 0, 1, and $-1$, respectively. The model has
superpotential
\eqn\supbare{W=\phi X\tilde\phi + M^2 X-{\lambda\over 3} X^3}
and $D$-term
\eqn\Dbare{D = e|\phi|^2-e|\tilde\phi|^2-r \ . }
So far the model has two parameters input by hand: $r$ and $M$.

Taking $\phi,\tilde\phi$ to be messengers, a SUSY breaking
configuration with $\langle X \rangle, \langle F_X \rangle \neq 0$
would transmit SUSY breaking to the Standard Model a la gauge
mediation. This model has such a minimum, as follows. The potential
energy of the model is
\eqn\Vtot{
V(\phi,\tilde\phi,X)
= |X|^2 (|\phi|^2+|\tilde\phi|^2) + |\phi\tilde\phi+M^2-\lambda X^2|^2
+ {1\over 2}(e|\phi|^2-e|\tilde\phi|^2-r)^2+\Delta V \ , }
where $\Delta V$ is the Coleman-Weinberg potential expanded about
the point of interest in field space.

To obtain the structure described above, let us expand the theory
about $\phi = \tilde\phi = 0$ and $X \approx M/\sqrt{\lambda}$.  We
assume $X^2 \gg eD$, and also take $eD \gg F_X \equiv M^2 - \lambda
X^2$; we will verify that the latter assumption is self-consistent at
the end.  With these hierarchies, the $\phi, \tilde\phi$ origin is
stable, with $m_\phi^2 = |X|^2+eD$ and $m_{\tilde\phi}^2 =
|X|^2-eD$. Setting $\phi = \tilde\phi = 0$, we find the following
potential for $X$:
\eqn\Vphi{V_{\rm eff}( X)=|M^2-\lambda X^2|^2+{1\over 2}D^2+\Delta V \
, }
where, at the present level, $D=r$ is an input constant.  In the
dynamical version to follow, we will render $D$ dynamically small in
the vacuum.

The Coleman-Weinberg potential $\Delta V$ is straightforward to
calculate here, particularly given $eD\gg F_X$. It is
\eqn\Vdelt{\Delta V( X)= {\rm Tr~ log} \bigl((| X|^2+p^2)^2-
e^2D^2\bigr)-
{\rm Tr~ log} \bigl(| X|^2+p^2\bigr)^2+{\cal O}(F_X^2) \ .}
Here the first term comes from the $\phi,\tilde\phi$ loops, and the
second comes from the fermion loops which must cancel the first term
up to the subdominant $F$-breaking effects.  Performing the
integration over momentum gives the result
\eqn\VdeltII{\Delta V( X)= {{e^2D^2}\over{16\pi^2}} {\rm log} \bigl(|
X|^2/M_*^2\bigr)+{\cal O}(F_X^2)+{\cal O}(D^4e^4/ X^4) \ . }

This potential \Vphi\ has extrema at
\eqn\solphi{
X_{\pm}^2={M^2\over{2\lambda}}\biggl(1\pm\sqrt{1-{{e^2D^2}\over{8\pi^2
M^4}}}\biggr) \ , }
of which $ X_+\equiv X_0$ is a metastable minimum.  In the regime
defined above, this yields
\eqn\results{ X_0^2\approx
{M^2\over\lambda}-{{e^2D^2}\over{32\pi^2\lambda M^2}} \ , \quad
F_X\approx {e^2D^2\over{32\pi^2 M^2}} \ . }
As a self consistency check, for $M\gg eD$, we have $F_X \ll eD$, as
assumed above in the calculation of the Coleman-Weinberg potential.

It is worth noting that the result \results\ for $F_ X$ follows from a
simple scaling argument, which could be useful in more complicated
examples.  Before including the $D$-term breaking effect and resulting
Coleman-Weinberg potential, the theory had a supersymmetric vacuum at
$ X=M/\sqrt{\lambda}$, with $X$ mass $m_X = M$. The perturbative
correction to the potential produces a tadpole $\del
\Delta V/\del X$ evaluated at $ X_0\sim M$, which shifts the field by
an amount $\Delta X\sim (\del \Delta V/\del X)/m_ X^2$. The resulting
$F$-term is then of order
\eqn\Festimate{F_ X\sim {\del F\over{\del X}}\Delta X \ ,}
which agrees with the solution \results\ in the present example.

Altogether, we have recovered the standard structure of gauge
mediation in a simple model of perturbative SUSY breaking.  In this
example, the messengers participate directly in the SUSY breaking
dynamics, in that their radiative effects generate the
Coleman-Weinberg potential.  Hence this constitutes a model of direct
mediation. So far we have two input parameters, $eD$ and $M$. The
former is the only very small input scale required in the model, and
we will render it dynamically small in the next subsection. Tying $M$
to a dynamical scale would be somewhat more complicated.

\subsec{Dynamical $D$}

To render $eD$ dynamically small, we first trade it for a
superpotential term using the original Fayet model.  Add two chiral
fields $a,\tilde a$ of charge $\pm 1$ under the $U(1)$ symmetry, and a
superpotential
\eqn\subaI{W_{a0}=m_{a0} a\tilde a \ .}
As above we will be interested in large $X$, where $\phi =
\tilde\phi = 0$.  In this regime, for $er\ge m_{a0}^2$, the
minimization in $a,\tilde a$ yields a vacuum
\eqn\avac{e|a|^2=r-m_{a0}^2/e \quad eD=m_{a0}^2 \ . }
Thus the input Fayet-Iliopoulos parameter $r$ itself can be of order
the large scale $M_*$, and the problem of obtaining a naturally
small $eD$ reduces to that of obtaining $m_{a0}$ dynamically.

This can be done as follows. First, note that the model would respect
a $\IZ_2$ $R$ symmetry under which $a,\tilde a$ are neutral (and under
which $ X$ is neutral, with $\phi\tilde\phi$ transforming
nontrivially), if $m_{a0}, M^2,\lambda$ were replaced with a dynamical
operator which transforms nontrivially under the symmetry.  Introduce
a pure $SU(2)$ sector, with kinetic term $\int d^2\theta W_\alpha
W^\alpha$.  Here $W_\alpha W^\alpha$ transforms nontrivially under the
$\IZ_{2}$ $R$ symmetry so that this kinetic term is invariant under
the symmetry. Imposing this symmetry, we cannot write down a bare
$m_{a0}a\tilde a$ term, but we can write
\eqn\Wadyn{W_{a\Lambda}=a\tilde a W_\alpha W^\alpha/M_*^2 \propto
a\tilde a \Lambda^3/M_*^2 \ , }
which weakly couples the $SU(2)$ degrees of freedom to the
O'Raifeartaigh/Fayet SUSY breaking sector.  In the last step in
\Wadyn, we replaced $W_\alpha W^\alpha$ with its holomorphic VEV
$\Lambda^3$.  As in the O'Raifeartaigh case discussed above, it is
consistent to integrate out the Yang-Mills degrees of freedom, since
they couple weakly via $M_*$-suppressed couplings to the rest of the
theory.

By the same token, the above symmetry prevents the pure
superpotential $M^2 X -\lambda X^3/3$ from appearing, but this times
$W_\alpha W^\alpha/M_*^3$ can appear, along with an $M X^2$ term.
(Adding an additional symmetry-respecting term proportional to
$\phi\tilde\phi$ has no effect as it can be absorbed by a shift in
$X$.) This modification leaves fixed the scaling of the $X$ VEV
found above, $X_0\sim M$, and the scaling \Festimate\ of the
resulting $F$ term. Altogether this produces a theory in which the
small parameter $eD$ has been effectively replaced with $m_a^2\sim
\Lambda^6/M_*^4$. This leads to $F_ X \sim
\Lambda^{9}/M_*^7$.\foot{One could also consider the symmetry $X\to
-X$ under which $\phi\tilde\phi$ is invariant.}

This much is sufficient to obtain very high scale gauge mediation
naturally, with weak scale SUSY breaking obtained via the above
method for rendering $eD$ dynamically small, and with $M$ an order
of magnitude or two below $M_*\equiv M_{GUT}$ as the only input
parameter. If we take $M\sim 10^{-1}M_{GUT}$, then we obtain a high
scale gauge mediation model with a naturally small SSM gaugino mass
arising from the dynamically small $eD$ we obtained via the
retrofitting procedure.

\newsec{Discussion and Future Directions}

In this paper we combined simple ingredients in a straightforward
way to obtain SUSY breaking models with all hierarchically small
scales naturally explained dynamically.  This procedure of
retrofitting simple models can, of course, also be applied to more
intricate examples; for example one can similarly retrofit the model
of \ISS\ to render the input quark mass scale dynamically small, as
was done recently in a footnote in \SchmaltzQS.  In retrospect,
however, perhaps the simplest possibility for model building is to
obtain the small scale as a supersymmetric but dynamically small
VEV, while obtaining the breaking of SUSY the old fashioned way.

There are several future directions to pursue.  Here we focused on
perhaps the very simplest models of perturbative SUSY breaking, but
there are more general classes containing gauge fields for which one
can systematically analyze the vacuum structure and retrofitting. It
will also be interesting to investigate the realization of these
models in string compactifications.

It would also be interesting to investigate retrofitting models to
yield low scale messenger masses.  Gauge mediation models with
messenger masses below $\sim 10^7~{\rm GeV}$ have the desirable
feature that they do not require non-standard cosmology to avoid
overclosing the universe with gravitinos, and they predict the
spectacular prompt photon and multi-lepton collider signals usually
associated with gauge mediation.  Most direct gauge mediation models
discussed previously predict intermediate or high scale messenger
masses, in part because their extra particle content would otherwise
force couplings to Landau poles well below the GUT scale.  The
explicit examples of \S2.1 and \S3 also yielded intermediate and high
scale messenger masses.  However, as noted in
\S2.2, low scale models may be possible, especially given the
simplicity of the class of models discussed here.

Realistic application of these models requires an assessment of
their cosmological stability.  The metastable vacua themselves are
very long-lived, but whether the universe finds its way into them
cosmologically is an {\it a priori} separate question.  This is very
plausible given the symmetries governing our system \dinesymm.  It
is under investigation in a similar class of models along the lines
of \ISS\ in \Jay, and may be affected by the process described in
\trapping.

\medskip

\noindent{\bf Acknowledgments}

We fully acknowledge valuable discussions with T.~Banks, M. Drees,
K.~Intriligator, S.~Kachru, M.~Luty, J.~McGreevy, H. Murayama,
N.~Seiberg, Y.~Shadmi, and J.~Wacker.~ MD is supported in part by
the Department of Energy and a J.~S.~Guggenheim Fellowship, and
acknowledges the hospitality of the KITP where some of this work was
performed. The work of JLF is supported in part by NSF CAREER grant
No.~PHY-0239817, NASA Grant No.~NNG05GG44G, and the Alfred P.~Sloan
Foundation. ES is supported by the DOE under contract
DE-AC03-76SF00515 and by the NSF under contract 9870115.

\medskip

\listrefs

\end